# A Deep Learning Approach to Predicting Ventilator Parameters for Mechanically Ventilated Septic Patients

Zhijun Zeng*[1], Zhen Hou*[2], Ting Li*[3], Lei Deng[3], Jianguo Hou[4], Xinran Huang[5], Jun Li[6], Meirou Sun[4], Yunhan Wang[2], Qiyu Wu[4], Wenhao Zheng[7], Hua Jiang[#3], and Qi Wang[#8]

[1]School of Mathematics, Shanghai University of Finance and Econonmics, Shanghai, China,200433

[2]School of Mathematics, Nankai University, Tianjin, China 300071

[3]Institute for Emergency and Disaster Medicine, Sichuan Provincial People's Hospital, University of Electronic Science and Technology of China, Chengdu, China, 610072

[4]Beijing Computational Science Research Center, Beijing, China, 100931

[5]University of Glasgow, Glasgow G12 8QQ, United Kingdom

[6] School of Mathematics, Tianjin Normal University, Tianjin, China, 300387

[7]University of Electronic Science and Technology of China, Chengdu, China, 611731

[8] Department of Mathematics, University of South Carolina, Columbia, SC 29208, USA

*These authors contributed equally to this work as co–first authors.

#Corresponding author. Tel: 86-13980001701; Email: jianghua@uestc.edu.cn

#Corresponding author. Tel: 1-850-443-0294; Email: qwang@math.sc.edu.

**Author contributions:** The authors meet criteria for authorship as recommended by the International Committee of Medical Journal Editors. QW and HJ designed the study, supervised the research project, drafted and edited the manuscript. TL, LD and HJ conducted data acquisition and clinical management. QYW and XRH conducted statistical analyses and data preprocessing, WHZ, YHW, MRS, JGH and JL developed the LSTM method, ZJZ, ZH designed and implemented the LSTM algorithm, conducted the numerical computation and analysis, and provided initial draft of the method section. All authors contributed to the production of the final manuscript.

**Funding**：Research is partially supported by grant from NSFC (award NSAF-U1930402, 11971051). Hua Jiang's work is partially supported by Sichuan Department of Science and technology (2019YFS0534，2021YFH0109). Qi Wang's work is partially supported by National Science Foundation (award DMS-1815921, DMS-1954532 and OIA-1655740) and a GEAR award from SC EPSCoR/IDeA Program.

**Running title:** Predicting Ventilator Parameters for Septic Patients

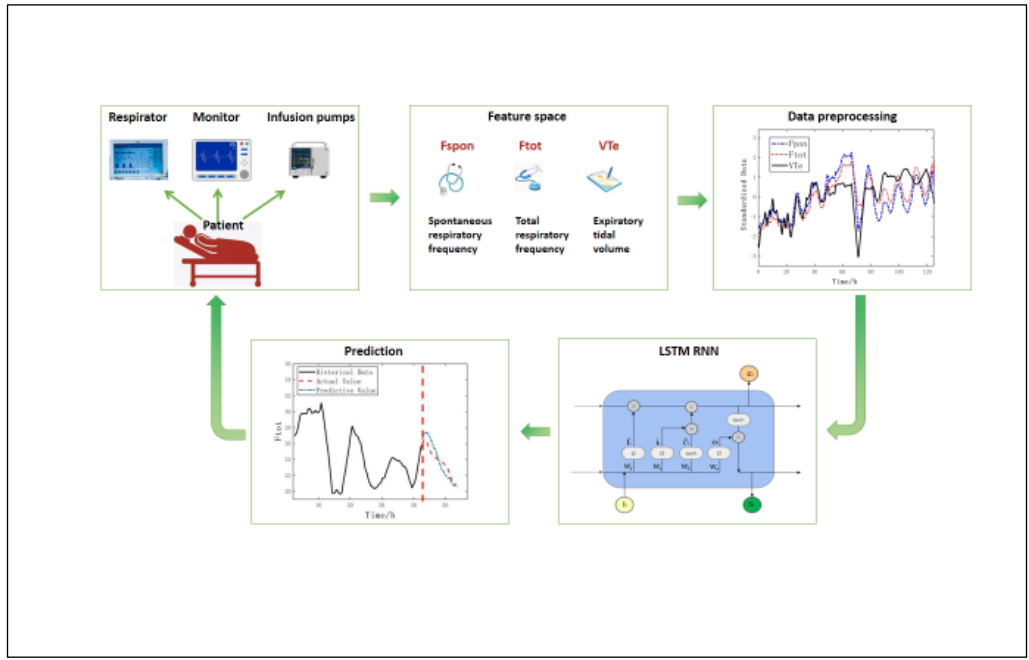

Graphical abstract

## Abstract

We develop a deep learning approach to predicting a set of ventilator parameters for a mechanically ventilated septic patient using a long and short term memory (LSTM) recurrent neural network (RNN) model. We focus on short-term predictions of a set of ventilator parameters for the septic patient in emergency intensive care unit (EICU). The short-term predictability of the model provides attending physicians with early warnings to make timely adjustment to the treatment of the patient in the EICU. The patient specific deep learning model can be trained on any given critically ill patient, making it an intelligent aide for physicians to use in emergent medical situations.

# 1 Introduction

Sepsis is a devastating clinical syndrome triggered by a maladjusted host response to infection that often leads to organ failure and ultimately death among patients admitted to the intensive care unit (ICU) [1]-[2]. Septic patients often develop the acute respiratory distress syndrome (ARDS), leading to the refractory hypoxemia and respiratory distress [3]. As a result, the patients' oxygenation indices decrease and they cannot breathe on their own so that they need mechanical ventilator assistance to breathe while closely monitored by physicians [4].

It is known that monitoring vital ventilator parameters can not only assist the treatment of ARDS patients, but also address issues triggered by the mechanical ventilation circulation line. The existing monitoring of a patient's ventilation parameters can only issue warning signals when critical values are crossed. When this happens, the patient condition might have deteriorated, leaving physicians few options to mitigate the situation. Therefore, how to dynamically monitor vital ventilator parameters and be able to issue advanced warnings so that physicians can take timely or even preemptive measures in the treatment becomes a pressing issue in clinical practices.

Given the wide range of types of septic patients and complex septic pathophysiology, the traditional clinical paradigm of randomized controlled trials using simple statistics can no longer address the clinical issue. One needs a new, early warning system to prognosticate patients' conditions and trend of their vital parameters. Deep learning can then be used to fill the gap.

Deep learning, combining neural networks with machine learning, has achieved a great deal of success in many fields today [5]. In life science, deep learning has been primarily used in imaging analysis and related diagnostic processes [5]-[7]. In medicine, evolution of a disease, patient recovery from an illness, medical treatment processes all generate huge amount of time series data. If the data could be harnessed fully and made intelligently use, one would be able to design better treatment pathways, make far more accurate diagnoses and even design effective preventive measures to keep people healthy [8].

The recurrent neural network (RNN) is a special type of neural networks suitable for time series data. To avoid potential vanishing or exploding of gradients in network training, one innovated the long short-term memory (LSTM) RNN [5][9], abbreviated as LSTM in this paper. LSTM adds various control gates to avoid the gradient diminishing and exploding issue during training, making it a capable deep learning tool to deal with time series data arising from a complex dynamical process.

In this study, we focus on three ventilator parameters of a septic patient: spontaneous respiratory frequency (Fspon), total respiratory frequency (Ftot) and expiratory tidal volume (Vte). The abnormal respiratory rate is a manifestation of the pulmonary dysfunction. The tidal volume is an important ventilator parameter for proper ventilation to take place. Equally important as VTe, inspiratory tidal volume (VTi) is the amount of air moved during the inspiration in the mechanical ventilator. Since VTi correlates strongly with VTe, we only couple VTe with Fspon and Ftot in this study. Notice that the goal of using a ventilator on a septic patient is to deliver a tidal volume large enough to maintain adequate ventilation but small enough to prevent lung injury. Hence, a dynamical system model that not only models the time series data coming out of the ventilator, but also predicts future trend and values would be extremely useful for physicians in ICUs.

## 2 Methods

### 2.1 Data acquisition and preprocessing

We briefly describe the septic patient whose data are used to develop the model. A young man in his thirties suddenly lost his consciousness and was admitted to the Sichuan Provincial People's Hospital at 01:00 on Nov 7, 2020. After his admission, emergency room physicians conducted biochemistry test and found his pH value at 7.02, blood glucose 54.3 mmol/L and there were no positive findings in head computerized tomography (CT) scan. He was diagnosed as type 1 diabetes, accompanied by ketoacidosis (KTA) and hyperosmolar coma. He was immediately sent to the emergency intensive care unit (EICU). Upon his admission to EICU, his vital signs were: temperature 35.7º C, pulse 140 bpm/min, respiratory rate 25 bpm/min, blood pressure 99/71 mmHg, APACHE II score 34 and SOFA score 14. Physicians further tested more inflammatory markers and the chest CT scan on the patient. The CT image indicated spread shadow and consolidation in his lower lobe in both lungs and the inflammatory markers showed severe infection (Procalcitonin (PCT): 52.74ng/ml, hypersensitive C-reactive protein (hsCRP): 87.85mg/ml, MONO CD 64 Index: 172.32). The evidence indicated that he was suffering from a septic shock. He received endotracheal intubation and was put on a mechanical ventilator. Intravenous insulin was bumped to alleviate the hyperglycemia; noradrenaline and fluid resuscitation were intravenously administered to keep his circulation system stable. After these interventions, his condition was stabilized for several hours. Soon after, the patient's oxygen index went down and venous-venous extracorporeal membrane oxygenation (V-V ECMO) was applied. The ECMO was removed on November 13 because of the patient's oxygen index and respiratory parameters were improved and the vital signs were also stabilized. But on November 16, the patient appeared to suffer an intestinal rupture; his family decided not to take any further interventions. He was pronounced died at 08:00 on November 16, 2020.

During his stay at EICU, his ventilator data were recorded and stored in the bedside server. For this study, we retrieved his time series data in 8 days from the server, covering the period from 15:31:28 on 11/07/2020 to 07:47:11 on 11/16/2020 and including the use of vasoactive drugs, fluid volume, and invasive/noninvasive physiological monitoring parameters. The four ventilator parameter data in Fspon, Ftot, VTi and VTe are included in the dataset. We conduct a correlation analysis among the four-parameter data and tabulate the result in Figure 1. Since the time series of VTi and VTe correlate strongly (correlation coefficient=94.6%), we decide to couple VTe only with Fspon and Ftot in this dynamical model. The study protocol for this patient has been approved by medical ethics committee of Sichuan Provincial People's Hospital (approval number:2021030), and consent informed acquired from the family of the patient.

As the parameter data retrieved from the ventilator record are recorded at unequal time intervals, we have to preprocess them for machine learning. These preprocessing procedures include 1).approximating the data using data fitting interpolations [10]-[11]; 2). re-sampling the interpolation functions to create three new time series for Fspon, Ftot and VTe with equal time intervals of length 60s, respectively; 3).averaging the time series using a sliding window of 400s to generate smoothed data from the time series obtained from step 2 [12]; 4). re-sampling smoothed datasets every other 8 points for all three parameters to create coarse-grained time series data. Note that the choice of the elapsed time step, 8, in coarse-graining is the result of extensive numerical experiments. The coarse-grained data in the last 5 days are plotted in Figure 2.

For the coarse-grained data, we standardize them to make them ready for machine learning by $\tilde{y}_i = \frac{y_i - \bar{y}_i}{\sigma_{y_i}}$, where $\{y_i\}$ is the generic notation for each of three coarse-grained time series in Fspon, Ftot and VTe, and $\bar{y}_i$ and $\sigma_{y_i}$ denote the mean value and the standard deviation of $\{y_i\}$, respectively.

## 2.2 LSTM RNNs

RNNs are often used to process time series data and learn underlying dynamics or patterns from the time series' historical data [5]. However, the traditional RNN has an inherent problem in gradient vanishing and explosion during training when applied to long time series. As the result, the LSTM RNN emerges as a solution to mitigate the problem. In LSTM RNNs, one adds a new state variable $C_t$ based on the hidden state variable $S_t$ and designs a set of gates to control the state of $C_t$. With $C_t$, one can implement the long and short term memory effect in the modified RNN to mitigate the potential gradient vanishing and exploding effect in machine learning.

The LSTM RNN with time step $\tau$ is summarized as follows, for $t = 1, \cdots, \tau$,

Forget – gate: $f_t = \sigma(W_f[S_{t-1}, I_t] + b_f)$,

Input – gate: $i_t = \sigma(W_i[S_{t-1}, I_t], b_i)$,

Cell – gate: $\tilde{c}_t = \tanh(W_{\tilde{c}}[S_{t-1}, I_t] + b_{\tilde{c}})$, (2.1)

Cell – state: $c_t = f_t \odot c_{t-1} + i_t \odot \tilde{c}_t$,

Output– gate: $o_t = \sigma(W_o[S_{t-1}, I_t] + b_o)$,

Hidden– state: $S_t = \tanh(c_t) \odot o_t$.

Here, $W$'s and $b$'s are the weights and biases of the neural network, and $\sigma$ is the sigmoid function. Figure 3 depicts the structure of the LSTM RNN in a computational graph.

## 2.3 Machine learning LSTM RNNs

We use PyTorch to train and test the LSTM model for the three ventilator parameters [13]. In order to predict future values of the parameters beyond a time marker $t_n$, we take BatchSize, N, as the largest time step of the measured data in the training set, i.e., we use the time series data in time interval $[t_{n-N}, t_n]$ to construct the training dataset for a given time marker $t_n$. To obtain an m-step prediction beyond $t_n$ using the training set and compare it with real world data (ground truth) in time interval $[t_{n+1}, t_{n+m}]$, we put the $m$ values in the time series data in $[t_{n+1}, t_{n+m}]$ from the measured data to the test set.

When we train the LSTM model using the training set in (Fspon,Ftot,Vte), we use all N sub time series data obtained by dropping the data from $t_{n-N}$ to $t_{n-1}$ incrementally. The resulting N sub-series are of size $N- k, k = 0, 1,\cdots ,N - 1$, respectively. For the sub-series of size less than N, we use zero-padding to fill zeros in front of the sub-series to make them series of length N. For every padded time-series, a corresponding target time-series is constructed by adding the real world (ground truth) value of the next moment and dropping the data at the beginning of the series so that the target time-series possesses the same length as its corresponding time series data. These dataset pairs are used in PyTorch to train the neural network parameters.

In testing, we firstly input the data in $[t_{n-N}, t_n]$ as the historical value, record the output cell state $c_{t_n+1}$, hidden state $S_{t_n+1}$ and use output $o_{t_n+1}$ as the predicted value in the next step. Then, we use these predicted values to obtain the second predicted value and state variables. This process is repeated until we reach the end of the testing set at $t_{n+m}$.

## 3  Results

We train two LSTM models to make predictions for future values, where Hidden Size represents the number of features in the hidden state *S*; BatchSize N denotes the maximum number of time steps before $t_n$ in the training set; Epoch is the stopping epoch in training of the LSTM model. The hyper-parameter values are obtained through extensive numerical experiments based on a synthetic dataset chosen from the original dataset. To make a short term prediction beyond given $t_n$, we use the models to predict the value of ventilator parameters up to $\frac{1}{3}$ length of the N time steps after $t_n$ in the following. For example, if N =300, we predict values in $[t_{n+1}, t_{n+100}]$ altogether. We compute errors in maximum norm $L_\infty$, mean square norm $L_2$ and mean absolute value norm $L_1$ between the predicted values and measured ones, respectively [5].

We choose two time markers ($t_n$) to divide the data into the training and testing set while training the two LSTM models with respect to their respective hyperparameters. We name the resulting models: model 1 and 2, whose hyperparameters are tabulated in Table 1. We then use both models to make predictions beyond their respective marker $t_n$ for ventilator parameters Fspon, Ftot and VTe, respectively. The results are summarized in Figure 4. The predictive errors in three different metrics: $L_1$, $L_2$ and $L_\infty$ norm are shown in Table 2, 3, 4, respectively.

The results show that both models have their predictive powers. Firstly, the models predict the trend of the time series in the next 5-10 hours in all results correctly (Figure 4). Secondly, the average predictive error in the $L_1$ and the $L_2$ norm are in fact small clinically, 3/4 of the results in relative errors $\leq$ 10% (Table 2). Examining closely in the $L_\infty$ norm, we notice that the predictions made by both models at time marker 1 not only capture the increasing/decreasing trend, but also the numerical values within the first hour in less than 11.5 % overall. Model 1 performs well even in the next 5 hours with the $L_\infty$ error less than 12.4%. Except for the predictions in Fspon values in the first three hours measured with errors around 12%, all the other predictions are made with errors less than 9%. At time marker 2, model 2 outperforms model 1 in all parameter predictions with $L_\infty$ errors less than 10% up to hour 4. Except for the first hour, model 1 shows consistent less accurate predictions than model 2 though.

Notice that the least accurate prediction lies in parameter Fspon in the $L_\infty$ norm and VTe measured in the two average norms. The normal value of VTe for a patient is between 400 and 500. So we are concerned with whether the patient's VTe value falls outside the normal range in the near future. At time marker 1, both models predict the value of VTe stays under 500 in the next 10 hours while at time marker 2, both models predict the value would be below 400 in the next 2 hours which makes perfect sense clinically. We note that although numerical values of the relative errors appear to be large in a few cases, the trend of the time series is predicted correctly in the next a couple of hours.

In summary, both model 1 and 2 can make decent predictions on multiple time series in short time. Model 2 predicts better on more ventilator parameters than model 1 does in short time measured in the $L_1$ and $L_2$ norm. But, in a longer time period, model 1 outperforms model 2 in the average norms. Through further analyses, we notice that this is because the batch size used in model 2 is larger than that in model 1. Hence, model 2 produces a better prediction for ventilator parameters that vary in long time than in short time in average norms. In the meantime, model 2 tends to overlook details in short time periods, thus yields a worse prediction over the ventilator parameters that vary significantly in short time than model 1. Measured in the $L_\infty$ norm, model 1 outperforms model 2 in the first dataset with time marker 1 while the performance is completely reversed in the second dataset. If we were to choose the proper model based on the $L_\infty$ norm, we would choose model 1 for the first dataset and model 2 for the second one. In practice, the optimal values of the hyperparameters have to be chosen through a large number of numerical experiments.

## 4  Discussion

Clinically, for patients receiving mechanical ventilation, physicians can adjust ventilator parameter settings when the patient's respiratory parameters become poor or unstable. When there are too many adjustable parameters associated with complex consequences of respiratory, circulation and metabolism systems, only well-trained physicians may have insights into what can be learned from the parameter settings in order to obtain a positive consequence related to specific ventilation parameter patterns. Hence, a useful intelligence aide for analyzing and forecasting the ventilator parameter patterns and underlying dynamics would lend physicians much needed helping hands in the midst of complex clinical situations.

In this study, we develop LSTM models to describe time-dependent dynamics in three major ventilator parameters: Fspon, Ftot and VTe of a septic patient. Fspon reflects spontaneous breathing of a patient. In different stage of sepsis and its treatment, Fspon values can vary from near zero to faster than 20 times/min. A smaller value may indicate the effect of deep analgesia and sedation. When the organ function improves and if Fspon is slower than 12 times/min, it suggests that spontaneous respiration is still poor, which may be related to deep analgesic sedation, un-metabolized drugs, weak breathing muscles and ventilator dependence. If it is faster than 20 times/min, it may be related to lung infection, poor ARDS control, poor lung compliance, or other organ functions and internal environmental factors. All the above situations require active monitoring and targeted treatment.

In the four predicted results of Fspon shown in the first column in Figure 4, the measured values and the predicted values are all under 12 times/min within the first hour after the time markers despite of some overshoots with errors consistently less than 17%. In longer time periods beyond the time markers, some predictions may be off by a larger percentage (Figure 4-j) despite that the trend is predicted correctly. If we examine the use of model 1 in the first dataset and model 2 in the second dataset, the $L_\infty$ error is consistently less than 12.32% up to the fourth hour.

Ftot is the sum of Fmand and Fspon. The setting of Fmand is determined according to the selected breathing mode and target minute ventilation and PCO2 level. Adult Fmand is set at 10 to 16 beats per minute. The LSTM model prediction of the Ftot values is consistently under 10% measured in the average norms or 16.6% in $L_\infty$ norm even in longer time periods shown in the second column in Figure 4.

The tidal volume (VT) includes VTi and VTe. The VTe is the actual tidal volume value of the patient exhales via the ventilator. In general, the monitored VTe and VTi are essentially the same, confirmed by the correlation analysis in Figure 1. The tidal volume is mainly affected by the preset pressure, resistance and compliance of the respiratory system, and its setting should ensure sufficient gas exchange. Usually 8 to 10 ml/kg is selected according to an ideal body weight. For patients with ARDS lung hyper protective ventilation strategy, physicians choose 4 to 6 ml/kg. The model predictions for the first hour are satisfactory. However, the prediction given by model 1 for time marker 2 shows a poor relative error measured in the $L_\infty$ norm beyond the first hour (Figure 4-i and Table 4). The general trend beyond the time markers is correct though, which offers some useful clinical significance nonetheless.

In 2016, surviving sepsis campaign expert group suggested that early detection and early treatment are needed to save sepsis patients [14]. There have been a number of studies focused on machine learning for sepsis from early screen to diagnostics [15]-[16]. However, most of the studies have yet made any noticeable clinical impact. There are several apparent reasons for this. Firstly, it is common that many studies used single-pass data for modeling [17]-[20]. Secondly, these methods rely on static data to make future predictions, have no bearing on dynamics of the diseases and therefore have resulted in less impressive clinical progresses. Obviously, a patient's disease progression is not only related to the severity when him/her admitted to ICU, but also to the dynamics of the disease and the treatment given. It clearly indicates that one needs to model dynamics of the impact factors. Thirdly, these studies used open source database data accompanied by confounding factors to build predictive models without external validation [21]-[22].

Our study is dynamical based on patient-specific time series data. Compared with traditional predictive models, our predictive models rule out the heterogeneity that varies from patient to patient to focus on prognosticating the result on the patient's specific time series data. These are the factors that many previous studies have overlooked. As the result, we can make a fairly accurate prediction for the first hour, which is very useful for clinicians in ICU. Besides, the patient specific model can be readily translated to other patients via transferred learning.

There are some limitations in the model: firstly, models can capture trends in important variables, but is less robust in predicting accurate values sometimes, especially, in longer time periods. Either more clinical variables need to be augmented to the dynamical system in order to describe its dynamics more accurately, or more extensive training needs to be done to train the model parameters, including the choice of hyperparameters. Both of them will the focus of our next study. In the future, we expect that this individualized predictive model is not limited to the prediction of important clinical parameters, but can also be applied to predicting the length of hospital stay, ICU demand, and risk of death.

## 5 Conclusion

We develop a LSTM RNN based modeling framework to model the ventilator time series data acquired from a critically ill septic patient in ICU. We demonstrate that the LSTM model can be used to make a qualitative prediction on the future time series data as well as their trend on three selected ventilator parameters in a clinically significant time frame. In most cases, a quantitative prediction can be achieved as well. This deep learning model provides a useful, intelligent aide for physicians to care for septic patients and responds effectively to precipitated medical events in ICUs. The patient specific model can be readily calibrated on any patient with analogous time series data.

**Word counts: 3496**

Table 1: Hyperparameters in the LSTM RNN.

| Hyperparameters | Hidden Size | Batch Size | Epoch | Input Size |
|---|---|---|---|---|
| Model 1 | 8 | 120 | 5000 | 3 |
| Model 2 | 14 | 300 | 3500 | 3 |

Table 2: Errors measured in the $L_1$ and $L_2$ norm.

| Time marker | Items | Model 1 $L_1$ | Model 1 $L_2$ | Model 2 $L_1$ | Model 2 $L_2$ |
|---|---|---|---|---|---|
| 1 | Fspon | 0.7616(8.31%) | 0.8693(12.82%) | **1.2474(13.6%)*** | 1.5031(17.83%) |
| 1 | Ftot | 0.8159(3.43%) | 0.8965(6.63%) | 0.9894(3.8%) | 1.1015(4.04%) |
| 1 | VTe | 10.1148(2.21%) | 11.9187(4.40%) | 8.8486(1.9%) | 9.9511(5.46%) |
| 2 | Fspon | **2.0056(25.33%)*** | 2.2787(22.94%) | 0.6085(7.68%) | 0.7065(7.06%) |
| 2 | Ftot | 1.9741(9.92%) | 2.2616(10.38%) | 0.5944(2.99%) | 0.6969(3.21%) |
| 2 | VTe | **73.7988(23.48%)*** | 89.606(23.09%) | 13.67(4.35%) | 16.9138(5.24%) |

* Although the error seems large for some predictions, the models capture the trend in the next a few clinically significant hours in the EICU setting.

**Correlations**

| | | ftot | VTe | VTi | fspn |
|---|---|---|---|---|---|
| ftot | Pearson Correlation | 1 | -.051** | -.221** | .859** |
| | Sig. (2-tailed) | | .000 | .000 | .000 |
| | N | 16109 | 16109 | 16109 | 16109 |
| VTe | Pearson Correlation | -.051** | 1 | .946** | -.283** |
| | Sig. (2-tailed) | .000 | | .000 | .000 |
| | N | 16109 | 16109 | 16109 | 16109 |
| VTi | Pearson Correlation | -.221** | .946** | 1 | -.439** |
| | Sig. (2-tailed) | .000 | .000 | | .000 |
| | N | 16109 | 16109 | 16109 | 16109 |
| fspn | Pearson Correlation | .859** | -.283** | -.439** | 1 |
| | Sig. (2-tailed) | .000 | .000 | .000 | |
| | N | 16109 | 16109 | 16109 | 16109 |

**. Correlation is significant at the 0.01 level (2-tailed).

*

Figure 1: The correlation coefficients of the four ventilator parameters in the data.

Table 3: $L_\infty$ errors of both models for the first dataset.

| Time marker1 | $L_\infty$ error | | | Relative error | | |
|---|---|---|---|---|---|---|
| Model 1 | FSPON | FTOT | VTE | FSPON | FTOT | VTE |
| Hour 1 | 1.297157 | 1.138645 | 4.980713 | 0.114451699 | 0.041525 | 0.010994 |
| Hour 2 | 1.373423 | 1.138645 | 4.781769 | 0.121180795 | 0.041525 | 0.010557 |
| Hour 3 | 1.365078 | 0.850052 | 10.38965 | 0.123115525 | 0.031286 | 0.022519 |
| Hour 4 | 0.8915 | 0.740179 | 19.58765 | 0.086496808 | 0.028097 | 0.041093 |
| Hour 5 | 0.528788 | 1.202818 | 19.79672 | 0.056614466 | 0.047446 | 0.040657 |
| Model 2 | | | | | | |
| Hour 1 | 1.276876 | 0.970741 | 12.14362 | 0.112662263 | 0.035402 | 0.026805 |
| Hour 2 | 2.381243 | 1.509581 | 11.60818 | **0.210103497** | 0.055053 | 0.025628 |
| Hour 3 | 2.880542 | 1.708344 | 12.29834 | **0.259794295** | 0.062876 | 0.026656 |
| Hour 4 | 2.854746 | 1.585318 | 12.29834 | **0.276978791** | 0.060177 | 0.025801 |
| Hour 5 | 2.36772 | 1.067636 | 7.078522 | **0.25349915** | 0.042114 | **0.014537** |

Table 4: $L_\infty$ errors of both models for the second dataset.

| Time marker2 | $L_\infty$ error | | | Relative error | | |
|---|---|---|---|---|---|---|
| Model 1 | FSPON | FTOT | VTE | FSPON | FTOT | VTE |
| Hour 1 | 1.977986 | 1.945263 | 59.33023 | 0.169423 | 0.082370251 | 0.136243 |
| Hour 2 | 3.176628 | 3.223448 | 142.5279 | **0.267858** | 0.135498613 | **0.334201** |
| Hour 3 | 3.746116 | 3.653992 | 150.3851 | **0.367226** | 0.165281177 | **0.373825** |
| Hour 4 | 2.224755 | 2.342337 | 86.02466 | **0.258161** | 0.113716684 | **0.242696** |
| Hour 5 | 2.631479 | 2.699215 | 86.62083 | **0.350975** | 0.138491243 | **0.266607** |
| Model 2 | | | | | | |
| Hour 1 | 1.164195 | 1.190359 | 10.10464 | 0.099718519 | 0.050405 | 0.023204 |
| Hour 2 | 1.201909 | 1.217506 | 10.405 | 0.10134694 | 0.051178 | 0.024398 |
| Hour 3 | 0.838879 | 0.737265 | 22.3103 | 0.082233958 | 0.033349 | 0.055459 |
| Hour 4 | 0.603168 | 0.549547 | 33.37863 | 0.069991775 | 0.02668 | 0.094169 |
| Hour 5 | 2.211946 | 2.273077 | 74.81775 | **0.295019269** | 0.116627 | **0.230278** |

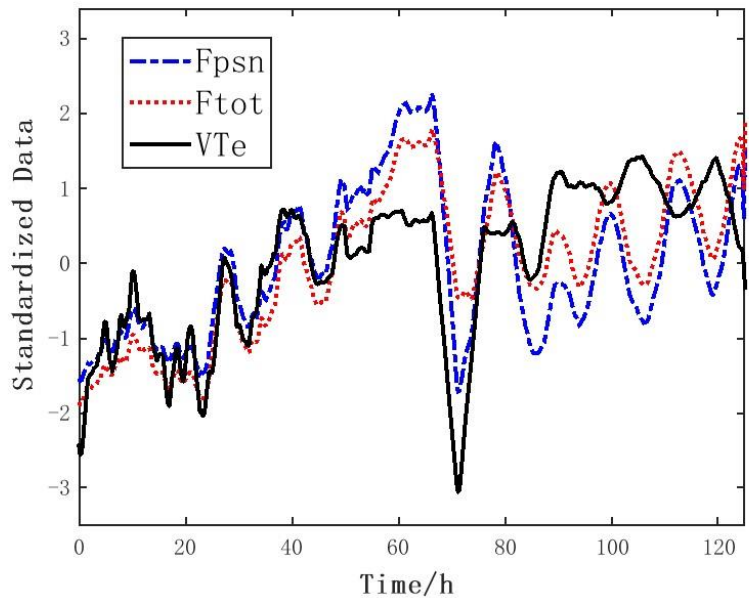

Figure 2: The Standardized Data in 5 days of the three ventilator parameters.

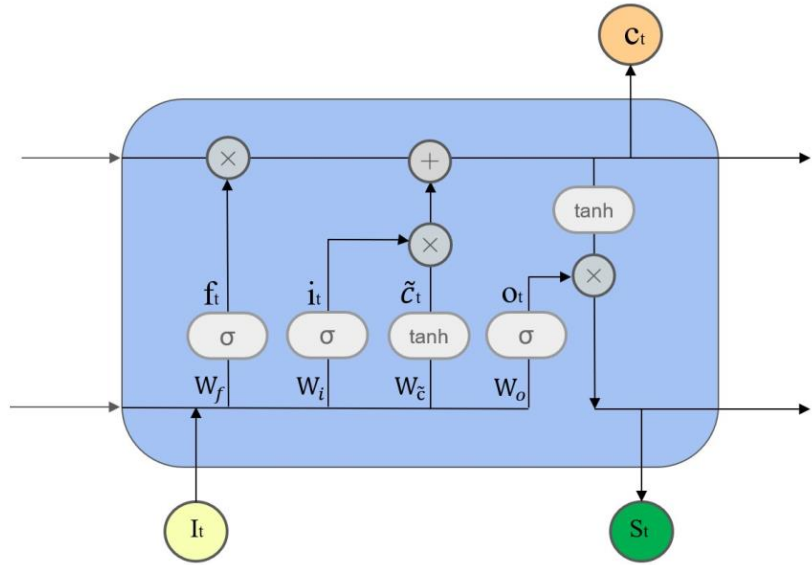

Figure 3: The structure of an LSTM RNN in its computational graph for $t = 0, \cdots, \tau$.

(a) Model 1 prediction for Fspon beyond time marker 1

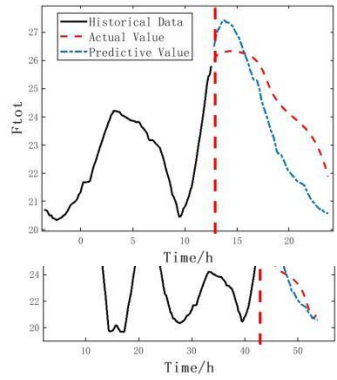

(b) Model 1 prediction for Ftot beyond time marker 1

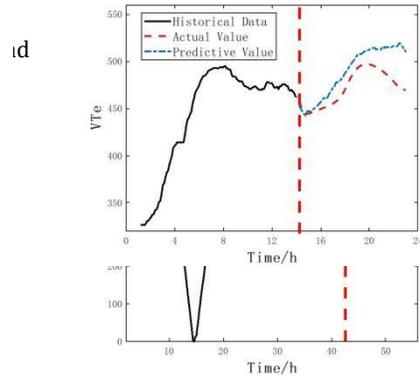

(c) Model 1 prediction for VTe beyond time marker 1

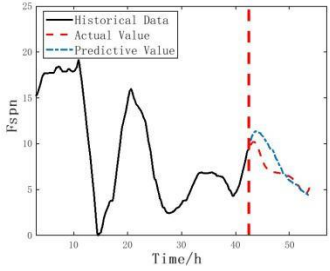

(d) Model 2 prediction for Fspon beyond time marker 1

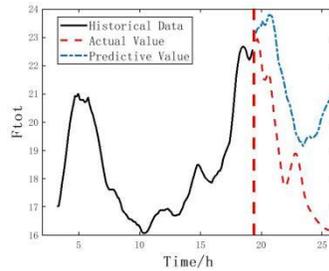

(e) Model 2 prediction for Ftot beyond time marker 1

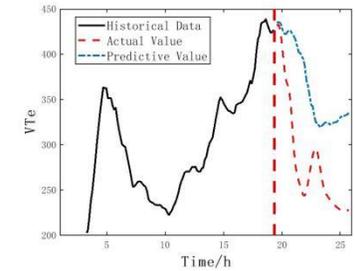

(f) Model 2 prediction for VTe beyond time marker 1

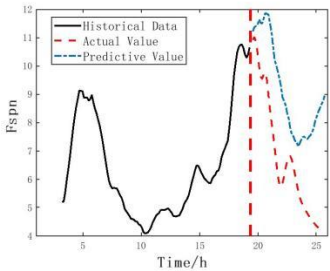

(g) Model 1 prediction for Fspon beyond time marker 2

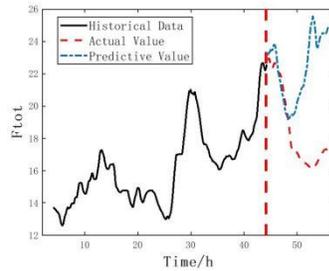

(h) Model 1 prediction for Ftot beyond time marker 2

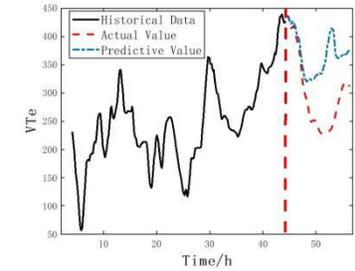

(i) Model 1 prediction for VTe beyond time marker 2

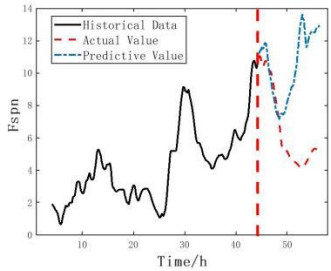

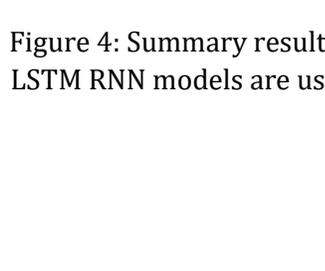

(j) Model 2 prediction for Fspon beyond time marker 2

(k) Model 2 prediction for Ftot beyond time marker 2

(l) Model 2 prediction for VTe beyond time marker 2

Figure 4: Summary results of the models. The vertical line sets the time markers, beyond which the LSTM RNN models are used to make predictions.